\documentclass[11pt]{scrartcl} %{article}
\ifx\pdfminorversion\undefined
\else
\fi

\usepackage{fullpage,epsfig}
\usepackage{comment}
\usepackage{xcolor}
\usepackage{longtable}
\usepackage{multirow,array}
\usepackage{authblk}
\usepackage{parskip}
\PassOptionsToPackage{hyphens}{url}
\usepackage{hyperref}
\usepackage{color,xspace}
\usepackage{cancel}

\usepackage{enumitem}

\newcommand{\eg}{\emph{e.g.,}\xspace}
\newcommand{\Eg}{\emph{E.g.,}\xspace}
\newcommand{\ie}{\emph{i.e.,}\xspace}

\newcommand{\remove}[1]{}

\title{Privacy-preserving transactive energy systems: Key topics and open research challenges\\
\Large{A formal note of discussion sessions from 2023 PriTEM workshop}}

\author{Daniel Gerbi Duguma, Juliana Zhang, Meysam Aboutalebi, Shiliang Zhang, Catherine Banet, Cato Bjørkli, Chinmayi Baramashetru, Frank Eliassen, Hui Zhang, Jonathan Muringani, Josef Noll, Knut Inge Fostervold, Lars Böcker, Lee Andrew Bygrave, Matin Bagherpour, Maunya Doroudi Moghadam, Olaf Owe, Poushali Sengupta, Roman Vitenberg, Sabita Maharjan, Thiago Garrett, Yushuai Li, Zhengyu Shan}

\affil{University of Oslo, Oslo, Norway\footnote{E-mail: \{daniegd, j.j.y.zhang, meysam.aboutalebi, shilianz, catherine.banet, cato.bjorkli, cpbarama, frank, huizhang, jonathan.muringani, josef.noll, k.i.fostervold, larsbock, l.a.bygrave, matin.bagherpour, m.d.moghadam, olaf, poushals, romanvi, sabita, thiagoga, yushual, zhengyus\}@uio.no}}%\\The authors are ranked in alphabetical order}}
\begin{document}
\maketitle

\begin{abstract}
This manuscript aims to formalize and conclude the discussions initiated during the PriTEM workshop 22-23 March 2023\footnote{\url{https://www.mn.uio.no/ifi/english/research/projects/pritem/events/conferences/workshop202303.html}, accessed %ible on 
Dec. 6, 2023.}. We present important ideas and discussion topics in the context of transactive energy systems. Moreover, the conclusions from the discussions articulate potential aspects to be explored in future studies on transactive energy management. Particularly, these 
conclusions cover research topics in energy technology and energy informatics, energy law, data law, energy market and socio-psychology that are relevant to the seamless integration of renewable energy resources and the transactive energy systems-in smart microgrids-focusing on distributed frameworks such as peer-to-peer (P2P) energy trading. We clarify issues, identify barriers, and suggest possible solutions to open questions in %those 
diversified topics, such as %e.g., 
block-chain interoperability, consumer privacy and data sharing, and participation incentivization. Furthermore, we also elaborate challenges associated with cross-disciplinary collaboration and coordination for transactive energy systems, and enumerate the lessons learned from our work so far.
\end{abstract}

\begin{comment}
    Topics for this manuscript
    1. Collaboration and knowledge exchanging & integration across multi disciplinaries  (Juliana)
    2. Privacy preservation and data utilization (Daniel)
    3. Barriers in transactive energy practice (Meysam)
    4. Break down of big energy transition goals and pathway for this transition (Shiliang)
    5. Block-chain related issues (Daniel)
    6. Socio-psychological issues (Juliana)
    7. Regulation and policy issues (Shiliang)
    8. Market and incentive issues (Meysam) % note on Friday Nov.\ 3, 2023: please consider to draw figure(s) on how the p2p energy trading happens in a local market with an aggregator -- this can serve as a common understading of our work.
\end{comment}

\section{Market, regulation, and policy for transactive energy systems}

\subsection{Transactive energy and peer-to-peer energy trading}

In energy informatics, the concept of transactive energy is well developed by the GridWise Architecture Council as: ``\textit{a system of economic and control mechanisms that allows the dynamic balance of supply and demand across the entire electrical infrastructure using value as a key operational parameter}"~\cite{osti_1123244}. Under the definition above, local energy markets can play a key role in implementing transactive energy systems, and in facilitating the mechanism to maintain the balance between energy generation and consumption. 

One innovative approach that empowers transactive energy activities is peer-to-peer (P2P) energy trading, which promotes a shared economy within local neighborhoods. P2P energy trading means direct energy exchange between individual consumers, often enabled by digital platforms~\cite{EES-031}. This model allows energy producers/prosumers, such as solar panel owners, to sell their surplus energy to other consumers in need. In that way, P2P energy trading empowers prosumers to actively participate and contribute as key stakeholders in realizing decentralized and distributed energy system frameworks while also optimally managing their energy consumption and production from various distributed energy resources~\cite{8582548}.

Due to its potential advantages, P2P energy trading is considered as an appealing alternative to traditional market structures. The European Commission has recognized its potential and prioritized it in the legal roadmap outlined in the Clean Energy Package~\cite{ 10.12688/openreseurope.15282.1}. P2P energy trading offers profits and benefits such as increased savings and greater autonomy for participants. The gained value depends on the grid tariff and market design. Typically, the price for P2P energy trading ranges between Feed-in Tariffs (FiT) and spot prices. However, the business model of P2P energy trading can vary depending on the market mechanism employed. We view the decentralized mechanism (full P2P market) and distributed mechanism (Hybrid P2P market) as the most viable alternatives to be developed and deployed. We consider such mechanisms well-suited for implementing P2P energy trading. When looking closely at 
the differences of these mechanisms,
%those mechanisms' differences, 
it is possible to assess their importance from several perspectives, including %. These included 
privacy, autonomy, scalability, uncertainty associated with renewable power generation, data sharing and data security, and power system operation~\cite{ZHOU2020739}. %The comparison considered various aspects to understand the advantages and disadvantages of each mechanism in the PriTEM project. It was suggested that both models could potentially be developed within the context of the project, enabling a detailed comparison of results using the same case study. 

Data flow plays a crucial role in the effective management of P2P and other forms of transactive energy trading. \Eg P2P energy trading relies on essential participant data, including asset specifications and consumption/production details. Such insights enable the local market to allocate the quantity and price of energy to be traded among participants through negotiation, clearance, and settlement processes. As P2P energy trading becomes more popular, there is also growing awareness and concern regarding data-sharing and privacy preservation. Such a situation is intensified as the collected participant data can disclose sensitive personal information, such as energy usage patterns, which can compromise home security or be leveraged by advertisers. Therefore, privacy and data security are critical to establish trust among participants and ultimately contribute to the real-world development of transactive energy trading frameworks.

The state-of-the-art research and ongoing pilot projects in P2P energy trading reveals significant challenges. These challenges include power system reliability, privacy preservation, data security, and trust enhancement. Addressing these challenges requires a comprehensive, holistic and cross-disciplinary approach involving technical, legislative, and socio-psychological dimensions to realize local energy markets in commercial scale- with seamless integration of renewable energy resources.

\subsection{Incentive, regulation, and policy issues}

\subsubsection{Incentives}

Incentive mechanisms are crucial in the development of transactive energy systems. Various incentivizing policies and contractual agreements have emerged to accelerate the investments for the integration of renewable energy resources in to the smart grid. These mechanisms aim to ensure the profitability of renewable energy generation, and guarantee the purchase of electricity produced by renewable resources through long-term contracts. Examples of such mechanisms include Net Metering~\cite{proceedings2231472}, Power Purchase Agreements (PPA)~\cite{Chapter12Longtermpowerpurchaseagreementsthefactorsthatinfluencecontractdesign}, and Feed-in Tariff (FiT)~\cite{POULLIKKAS20131}. These schemes create a stable economic framework that incentivizes the development and integration of renewable energy resources.

However, governmental subsidies for renewable energy tends to decline, and some long-term incentives are becoming less attractive-thus failing to effectively incentivize the public to invest in renewable power generation. In response to this challenge, innovative schemes are emerging. \eg in Norway, the concept of solar bank has been introduced, which offers seasonal storage solutions for the solar energy produced\footnote{More information available at: \url{https://midtenergi.no/solkonto/}, accessed %ible on 
Dec.\ 6, 2023}. These novel incentives aim to address the limitations of conventional incentivizing mechanisms and provide new opportunities for individuals to invest in and benefit from renewable energy generation. 

%As the permeation of renewable and storage systems in the residential sector is limited in Norway, it is of critical importance to come up with incentivizing policies to motivate consumers to install more distributed energy resources and possibly further to participate in P2P energy market. We envision that P2P energy trading can be a future market format that fully encourages consumer engagement and empowers user participation. While there are several pilot energy projects that enable P2P energy trading\footnote{See Smartly: \url{https://www.smartly.no} or Power2peer: \url{https://power2peer.com}, accessed Nov.\ 24, 2023.}, it is yet not sufficiently investigated and there is a lack of a structured policy framework for the particular case of the P2P energy market in Norway.

\subsubsection{Policy for renewable energy integration and novel market format}

The development of renewable energy resources in the end-use energy sector is supported by various schemes worldwide, encouraging more renewable energy installation, particularly solar PV in households~\cite{en15031229}. However, when looking at conditions in different countries, local and diversified issues can exist. Taking Norway as an example, Norway's consistent reliance on hydropower and its geographical location have resulted in slower progress in renewable developments, compared to countries like Germany. %Recently, the increased electricity price, the green energy transition, and updated policies have led to a rise in small-scale renewable investments.

Despite recent increases in renewable energy resources, there is still a lack of structured policy frameworks for local energy communities, local energy markets, and P2P energy markets in many countries including Norway~\cite{MALDET2022131805}. Given the limited residential renewable power generation in Norway, it is crucial to design an incentivizing grid tariff and policy in order to encourage investments in renewable energy resources from the edge stakeholders such as households and communities and participation in the P2P energy market, potentially leading to a high-level user engagement and consumer empowerment.

\subsubsection{Data protection legislation}

The penetration of transactive energy systems entails the generating and transferring of fine-grained consumer data for operation and management purpose. Nevertheless, high-resolution data risks the disclosure of user's private life and leads to privacy issues. Data and privacy protection has gained awareness and has been recognized as a barrier to the acceptance of transactive energy techniques.

The protection of consumer data and privacy is an essential element in ensuring that personal information is collected and used in a transparent and accountable manner~\cite{cavoukian2010privacy}. It includes the implementation of appropriate safeguards to protect the privacy rights of individuals. %Legislative privacy is also affected by the privacy-transparency conundrum.
An important challenge in data protection and privacy lies in technical implementation~\cite{koops2014privacy}. \Eg laws regarding data breach notification often require systems to notify users when their personal data is compromised. However, it is not clear how these laws apply to distributed networks, such as blockchain-based transactive energy systems. Due to the data immutability in blockchain, compliance with data breach notification laws can be difficult. Therefore, appropriate measures and technologies should be in place to guarantee privacy and data protection in transactive energy systems.

\subsubsection{Regulation issues with blockchain}

It is anticipated that blockchain will potentially be one of the main technologies to facilitate transactive energy systems in future energy market. However, the decentralized nature of blockchain platforms raises questions regarding data ownership and management, particularly in the context of blockchain interoperability. In a decentralized environment, it is nontrivial to establish clear accountability for data handling under the ambiguity of who owns the network, who has processed what data, where, and when\footnote{\url{https://widgets.weforum.org/blockchain-toolkit/interoperability/index.html\#q01}, accessed  Dec.\ 6, 2023.}. These complexities are further amplified when interoperability issues are considered, as multiple blockchain platforms and diverse data governance frameworks are involved. Moreover, Interoperability solutions must navigate a complex web of regulations, standards, and guidelines, ensuring compliance with both domestic and international requirements-which currently do not provide tangible concrete specifications. %That is, regulatory requirements for energy data management and transaction settlements, including data ownership and data retention policies, differ across jurisdictions. 

% \section{\cancel{Block-chain related issues}}

\section{Socio-psychological perspectives toward transactive energy activities}
So far, most of the works on transactive energy systems focus on technological aspects. They aim to address issues, \eg optimization of the electrical power system, trading algorithms and platforms, integration of renewable energy sources to existing grids, flexibility management, and other challenges in implementing pilot energy projects. A literature review for P2P energy trading~\cite{soto2021peer} reveals the common research topics as (i) trading platform (ii) blockchain (iii) game theory (iv) simulation (v) optimization, and (vi) algorithms. While these topics are certainly critical to the realization of transactive energy systems, we need to take a more holistic approach that takes into consideration the user perspectives and socio-psychological contexts. Although approaches like game theory relate closely with motivational psychology, the main focus in the state of the art is on mathematics, rather than on the psychological aspects of decision-making. Furthermore, energy is an integral part of every modern society, however, the research looking into the social contexts of energy management is insufficient. As such, there is a clear gap in transactive energy research regarding the human and social aspects of the equation.

\subsection{User perspectives and awareness}
Incorporating user perspectives in energy research is becoming increasingly vital as sustainable and efficient energy systems gain momentum. Understanding how users interact with and perceive energy technologies is crucial for the successful adoption and optimization of novel energy systems like P2P energy trading. There is a growing trend of user-centered design in energy technologies, such as various interactive energy management systems~\cite{dalen2017towards}\cite{ahram2010user}\cite{savchukuser} used to monitor and control energy consumption in modern homes. 

We briefly summarize the factors influencing the acceptance and adoption of P2P energy trading among prosumers, as highlighted in Table~\ref{tab:my_label}. We view the willingness to use green energy and motivation for cost-saving to be crucial among monetary and non-monetary factors.

\begin{table}[t!]
    \centering
    \begin{tabular}{ |p{4cm}|p{5.25cm}|p{5.25cm}|  }
 \hline
 \multirow{2}{0em}{References}    &\multicolumn{2}{c|}{Factors} \\
 \cline{2-3}
 & Monetary &Non-monetary\\
 \hline
 Culture, values, lifestyles, and power in energy futures: A critical peer-to-peer vision for renewable energy\cite{ruotsalainen2017culture}.
 & {\begin{itemize}[align=left, left=0em,labelsep=0.5em]
\item[-] Rising electricity prices
\item[-] Investments in local community
\item[-] Financial compensations
\end{itemize}}
 &{\begin{itemize}[align=left, left=0em,labelsep=0.5em]
\item[-]Concern for climate change
\item[-] Greater control and autonomy
\item[-] Strengthening of social cohesion
\end{itemize}}\\
\hline
Quantifying factors for participation in local electricity markets\cite{mengelkamp2018quantifying}.
 & {\begin{itemize}[align=left, left=0em,labelsep=0.5em]
 \item[-] Price consciousness
\end{itemize}}
 &{\begin{itemize}[align=left, left=0em,labelsep=0.5em]
\item[-] Technology affinity
\item[-] Importance of green products
\item[-] Community identity
\item[-] Regionality
\end{itemize}}\\
\hline
Keep it green, simple, and socially fair: A choice experiment on prosumers’ preferences for peer-to-peer electricity trading in the Netherlands\cite{georgarakis2021keep}.
 & {\begin{itemize}[align=left, left=0em,labelsep=0.5em]
\item[-] Selling prices
\end{itemize}}
 &{\begin{itemize}[align=left, left=0em,labelsep=0.5em]
\item[-] Reducing emissions
\item[-] Social connection
\item[-] Improved efficiency
\item[-] Self-sufficiency
\end{itemize}}\\
\hline
A Preference Analysis for a Peer-to-Peer (P2P) Electricity Trading Platform in South Korea\cite{li2022preference}.
 & {\begin{itemize}[align=left, left=0em,labelsep=0.5em]
    \item[-] Cost savings
    \end{itemize}}
 &{\begin{itemize}[align=left, left=0em,labelsep=0.5em]
        \item[-] Security
   \end{itemize}}\\
\hline
\end{tabular}
    \caption{The influencing factors to participate in peer-to-peer energy trading from  the literature review.}
    \label{tab:my_label}
\end{table}

Despite the %those 
identified factors, there have been limited studies conducted to formulate these factors into the design of P2P energy market. To narrow the gap, we suggest the methods like weighted optimization to prioritize and integrate the influencing factors in the business model of local energy markets.

There are also increased efforts to nurture understanding and improve user awareness and literacy, \eg toward energy consumption, sources, and impacts, which is crucial to promote energy-saving behaviors and the adoption of renewable energy technologies. The concepts of transactive energy and P2P energy trading are relatively new to ordinary energy users. As such, a clear understanding of the market structure among potential participants is essential, which might promote technology learning and active engagement in novel energy markets.

\subsection{Understanding the emerging role in energy market}

We observe that a new role called ``prosumer" emerges as distributed energy generation becomes an alternative to individuals and communities. A prosumer can both produce and consume energy, with capacity of local energy generation, \eg from solar panels. To fully harness the positive influences by prosumers, more research is needed to better understand their behaviors and factors affecting their decision-making. Prosumers are generally seen as active agents in local energy markets who can contribute to maintain the balance of demand and supply. Nevertheless, existing research has not fully analyzed the wide range of prosumer demographics, including different socio-economic, cultural, and age groups of the prosumers. %Neither have we formed sufficient understanding of how the prosumers perceive, consume, and produce energy. 
Consequently, we are still lagging in the formation of inclusive and effective energy policies and technologies, which is expected to play pivotal role in user engagement and sustainable behavioral change. 

A study on German household energy by Hackbarth and Löbbe~\cite{hackbarth2020attitudes} shows the willingness to participate in openness towards P2P energy trading is the precursor to the willingness to participate in P2P energy trading. The main motivation for the participation is revealed as the ability to share electricity and become more self-sufficient. Karami and Madlener's work~\cite{karami2022business}  shows %presents 
that cost savings and financial benefits are the main motivators among households in Germany. Despite the discrepancy on prosumer engagement motivators, we believe that the presence of monetary (\eg trading profits, tax saving) and non-monetary motivators (\eg sense of community, environment preservation) are necessary to ensure active prosumer engagement. However, there is lack of clarity on the effectiveness of these %those 
motivators and how they work in different contexts. Majority of the relevant research took background in developed nations like Germany, United States, Switzerland, \emph{etc.}, which raises the question of generalizability of the findings. There is also limited knowledge on the barriers preventing broader adoption of renewables and active user participation in the energy transition. %Narrowing the gap is essential as it is likely that we have to prioritize certain features over the others in the design and optimization of local energy market. 

\subsection{Social perspectives}
Energy as a commodity is deeply entrenched in every fabric of the society. Nonetheless, previous energy research is often decontextualized, despite the very goal of creating an equitable and sustainable society. Therefore, it is important to highlight the significance of cultural values and social norms that influence energy-related behaviors. Due attention should also be paid to the roles of energy governance and policy making that shape individual and community responses. Another recognized critical area of research is the intersection of energy systems with environmental justice. Such studies are anticipated to examine how the benefits and burdens of energy production and consumption are distributed across different communities, with a particular attention on marginalized and disadvantaged groups. Taking the example of implementing transactive energy management in less developed countries. While transactive energy management may offer benefits such as democratization of energy access and economic growth, there remain fundamental challenges, such as the absence of local supportive policies and stable internet connections that underpin the adoption of novel technologies. To fully reap the benefits of transactive energy management, implementation needs to be informed by knowledge that is localized and context-specific.

\section{Blockchain for transactive energy management}

% TODO:
% - Import references from Word
% - Include the other issues (interoperability and scalability) in the Introduction
% - Concisely rewrite 2.1

Transactive energy relates to energy trading and management by facilitating the integration of renewable energy resources into the existing power grid infrastructure and promoting market-based energy production values at the distribution level~\cite{huang2021review}. With prosumers as the active contributors, the power ecosystem fully supported by transactive energy has the potential to create a truly participatory and decentralized energy market. Distributed ledger technology is a key technology for such a decentralised/distributed energy market to securely store data.~\cite{andoni2019blockchain}.

Blockchain is the digital ledger technology that records transactions in a public or private peer-to-peer network~\cite{nakamoto2008bitcoin}. These transactions are permanently recorded as blocks and distributed among several machines. All blocks in the blockchain are connected to one another using cryptography, \eg via digital signatures and hashes~\cite{narayanan2016bitcoin}. Data in the blocks is tamper-proof, where the change of data in any block can be detected as the hash pointer to each block is stored in the next block as well-for all the blocks in the chain.
%The distributed ledger's transparency promotes trust and accountability among participants. Through a consensus mechanism, all parties verify the authenticity of each transaction before incorporating it into the blockchain's immutable record.
%When one transaction approaches the blockchain, all participants need to agree on the veracity of this transaction via a consensus process, and once such validity is established, the data will be recorded by the blockchain and it becomes immutable.

% While data immutability is beneficial in many ways, it also has drawbacks. The most significant downside is the disclosure of sensitive data such as energy usage patterns, location information and financial transactions, which breaches people’s privacy rights. Although most blockchains implement pseudonymous identities to protect the privacy of their users, it is still possible to link transactions to specific identities (such as through pattern matching) [Ref]. 
Smart contracts are fundamental component of blockchain technology.
While the concept of smart contract was first introduced by Nick Szabo~\cite{szabo1996smart} in 1994, their potential for general-purpose computing was only fully realized two decades later with the launch of Ethereum\footnote{\url{https://ethereum.org/en/smart-contracts/}}. A smart contract is a self-executing code on a blockchain that automates predefined agreements without the need for intermediaries.
Smart contracts facilitate automated energy trading and other energy management processes in blockchain-based transactive energy systems. They enable secure and transparent exchange of energy assets, \eg excess energy, electric vehicle charging, demand response~\cite{kirli2022smart}.

Although smart contracts and blockchain technology have the potential to transform the energy sector and promote transactive energy, they encounter various challenges that limit their widespread adoption. Below we contextualize and detail the recognized blockchain issues regarding transparency, privacy, interoperability, and scalability.

%\subsection{Transparency, privacy, and the paradox}
\subsection{The transparency-privacy dilemma}
\subsubsection{Transparency}
%Transparency refers to a system's ability to be open and accessible, allowing participants to view all relevant information and transactions, while also ensuring the integrity of these records. Transparency can be achieved by properties, \eg total traceability, an auditable ledger for transactions, data immutability, and decentralized network design~\cite{Dubey2022blockchain}. Blocks enable complete traceability as participants of the network can track the whole history of a transaction, from its inception to its current state. Transparency achieved through traceability can further improve participants' trust. Since the distributed ledger is replicated among network participants, all members have access to the transaction history and every member can audit the system unilaterally and transparently. The decentralized architecture of blockchain promotes transparency by allowing participants to unilaterally verify transactions and jointly engage in the decision-making process, therefore precludes any sole control of data by a single entity. Furthermore, the decentralized paradigm leverages consensus algorithms and cryptographic mechanisms to maintain the integrity of data stored in the blockchain. As a result, even if data is kept in each participant's local storage transparently, the data is assured to remain immutable.
Transparency refers to a system's ability to be open and accessible, ensuring participants' access to all critical information and transactions, as well as integrity of these transactions. Transparency can be achieved by properties, \eg total traceability, an auditable ledger for transactions, data immutability, and decentralized network design~\cite{Dubey2022blockchain}. Blockchain enables participants track the entire history of energy transaction through total traceability with immutable transaction records. The decentralized architecture of blockchain makes the distributed ledger to be replicated among network participants, where all members have access to the transaction history and every member can audit the system unilaterally and transparently. %Data immutability across the blockchain network guarantees energy data integrity. 
These transparency features make blockchain a key player in secure and trusted transactive energy systems.

\subsubsection{Privacy} 

Privacy indicates the ability to control the information that others know about a person and also the actions others can take based on that knowledge. Specifically, it is about retaining power over one's own identity by controlling  when and to what degree one's personal data can be accessed~\cite{peng2021privacy}. Privacy preservation is essential for safeguarding sensitive information such as individual identity. Privacy is an increasingly important consideration in data protection legislation~\cite{Chantal2022privacy}. Privacy preservation in the context of transactive energy systems promotes customer trust by preventing potential threats such as unwanted alteration of energy-related metadata.

The concept of privacy in blockchain is sometimes confused with anonymity and pseudonymity, even though there are significant differences among them. Anonymity means an individual’s identity is completely unknown, making it impossible to ascribe activities to a specific person~\cite{conklin2004principles}. An identity with anonymity cannot be identified, contacted, or tracked. Unlike anonymity, pseudonymous individual activities can be attributed to a specific identity. %Broadly speaking, privacy is complex and more generic notion, whereas anonymity and pseudonymity are separate techniques of hiding or disguising an individual's identity.

\subsubsection{The dilemma}
%In blockchain, two types of personal data might be recorded: public keys and transaction data. Such information can be protected via different cryptographic mechanisms like obfuscation, zero-knowledge proofs, and ring signatures~\cite{barsan2019public}. The usage of private blockchain composed of trusted organizations can provide an appropriate level of privacy~\cite{gai2019permissioned}.

%The processing of personal data, in Europe for instance, must adhere to data protection and privacy legislation such as the GDPR (general data protection regulation) \cite{gdpr}. However, unlike centralized systems, applying GDPR regulations on blockchain presents issues owing to compliance requirements.

While data immutability is one of the core requirements of transparency in blockchain, it poses a significant challenge to achieve the compliance of privacy regulations, \eg the General Data Protection Regulations (GDPR) in EU. For instance, GDPR stipulates the right to be forgotten, \ie individuals have the right to have their personal data erased under certain circumstances. Compliance with this criterion is complicated by the immutability of public blockchains, as removing data without affecting the entire blockchain is impossible. Thus, the privacy of the personal data stored on blockchain can conflict transparency.

Off-chain storage is a possible solution to this conflict. By keeping data separate, it is possible under off-chain storage to remove information when necessary, thus meeting the privacy requirement, \eg the right to be forgotten under GDPR. This is accomplished by storing only the data reference on the blockchain (a.k.a. on-chain) while the actual data is stored off-chain. %The data reference, which is often a hash digest, can uniquely represent the data, and if the data's content changes, the hash of the old data will no longer match the new hash, maintaining the integrity of the information. % COMMENT: the last sentence is not clear.

Off-chain storage can improve privacy in blockchain, however, it also introduces additional challenges~\cite{barsan2019public}. When data is stored off-chain, smart contracts can no longer access it directly. In that case, an interface is essential to connect the smart contract with the data. Nevertheless, introducing such an interface may expose users to extra privacy and security breaches. Moreover, when data from off-chain storage is removed, the blockchain reference hash will correspond to null or non-existent data. As the number of transactions and blocks increase, information on the blockchain that leads nowhere will accumulate over time, rendering the entire blockchain ineffectual.

%Apart from privacy regulations, the transparency-privacy dilemma of blockchain exists in financial laws, particularly concerning anti-money laundering (AML) and combating the financing of terrorism (CFT)~\cite{Karasek2021}. Due to the anonymous nature of financial transactions in most public blockchains, criminals may use this opportunity to conceal the origin of their funds and circumvent existing AML/CFT detection mechanisms. While the transparency and immutability of blockchain enables authorities to monitor fund flow and identify potential risks, it may disclose the privacy of individuals and organizations engaging in lawful transactions. Furthermore, because blockchain transactions are recorded on a public ledger, anyone with blockchain access can track transactions back to individual addresses and associate them to actual identities. This might jeopardize people' need for secrecy and anonymity, perhaps exposing them to threats such as identity theft, extortion, or profiling.

The discussed dilemmas call for a careful examination of the core of privacy and transparency, particularly when one is achieved sacrificing the other. The increased surveillance of individual transactions within the blockchain framework creates a delicate interaction that diminishes privacy in the presence of transparency. As transactions and various personal information are recorded in the public ledger, it becomes more challenging to achieve a good balance between personal privacy and societal transparency. As a result, it is critical to recognize when social welfare outweighs individual benefit in the trade-off between privacy and transparency. In this regard, deriving concrete metrics for trust and prosumer empowerment is crucial. Moreover, computation efficiency and energy efficiency of the technical solutions are important aspects for them to be of practical value.

\subsubsection{Concluding remarks}
In public blockchain, the privacy and transparency trade-off is about finding an appropriate balance between them, such that transactions become anonymous, while participants may still verify the information. These transactions are often maintained in an open environment where anybody may see and audit them, encouraging auditability and trust.  Transparent transactions, on the other hand, may result in the disclosure of personal data that was meant to be private. 

It has been observed that cryptographic approaches, such as zero-knowledge proofs and ring signatures, are used to ensure both transparency and privacy in public blockchains, while promoting trust and maintaining data security. Though openness and privacy appear to be diametrically opposed, the latter may not have to be compromised to maintain the former. That is, it is conceivable to preserve a reasonable amount of privacy while increasing openness, even if doing so comes with its own challenges.

\subsection{Blockchain interoperability}
Interoperability in blockchain visions the ability to seamlessly transfer both digital assets and transaction records across disparate networks, and eliminate the need for intermediaries exchanges. Interoperability technologies is supposed to facilitate a seamless and secure execution of smart contracts across diverse blockchain networks. In that way, data exchange at the foundational level is enabled by standardised data formats, and transaction data across systems becomes comprehensible to end-users\footnote{\url{https://towardsdatascience.com/blockchain-interoperability-33a1a55fe718}, accessed %ible on 
Dec.\ 6, 2023.}. However, realising the vision of seamless blockchain interoperability requires addressing several key challenges concerning standardisation, consensus protocols, smart contracts, and regulatory frameworks across jurisdictions.

\subsubsection{Standardization in blockchain interoperability}

In blockchain operations, functionalities, \eg sending tokens between participants, executing smart contracts, and ensuring data validity, are restricted to individual blockchains. This limitation articulates the problem of interoperability across different blockchain systems, while the situation is further deteriorated by the absence of standards. % TODO: add citation 
%The problem of blockchain interoperability, wherein functionalities like sending tokens between participants, executing smart contracts, and ensuring data validity are restricted to individual blockchains, is intensified by the absence of standards [3]. One critical reason for this limitation is the lack of established interoperability standards in existing protocols, hindering seamless interaction between different blockchains. 

While blockchain is a potential enabler for transactive energy trading, diverse data formats across blockchain networks hinder the interoperability. Though several organizations are actively involved in blockchain standardisation\footnote{\url{https://digital-strategy.ec.europa.eu/en/policies/blockchain-standards}, accessed %ible on 
Dec.\ 6, 2023.}, this work is still in its early stages. The lack of standardised data representations and inconsistencies in transaction formats pose challenges for aggregating, analysing, and processing energy trading information. Inconsistencies in metadata and communication protocols further restrict cross-platform data exchange, impeding real-time trading and market insights. Standardization initiatives are crucial to address these interoperability hurdles and facilitate efficient energy transaction management.

%Blockchain standardisation is still in its early stages. Several organizations are actively involved in this journey. European standardization bodies such as the European Telecommunications Standards Institute (ETSI), the European Committee for Standardization (CEN), and the European Committee for Electrotechnical Standardization (CENELEC) are working through their respective Joint Technical Committee 19 (JTC19) to develop common standards for blockchain technology~\cite{BCStandard}.

\subsubsection{Consensus mechanisms in blockchain}
The transaction processing speeds vary across blockchain networks, primarily due to the different consensus mechanisms employed in the blockchain. The choice of a consensus mechanism, such as Proof-of-Work (PoW), Proof-of-Stake (PoS), or Practical Byzantine Fault Tolerant (PBFT) protocols, significantly affects the security, transaction speed, and scalability of a blockchain network\footnote{\url{https://www.nec.com/en/global/insights/article/2020022520/index.html}, accessed Dec.\ 6, 2023.}. 

There is a heterogeneity of consensus mechanisms employed across blockchain networks in the context of energy transactions. This divergence can disrupt real-time energy trading, as transaction settlements may take longer on slower blockchains. Moreover, the combined effects of inconsistent transaction speeds and cross-platform discrepancies can undermine market efficiency in blockchain-based energy trading.

\subsubsection{Smart contract issues}
The diversity of smart contract programming languages and execution environments across blockchain platforms presents a significant obstacle to interoperability in blockchain-based energy trading. Programming for smart contracts differs across blockchain platforms, ranging from the Turing-incomplete Bitcoin script to the Turing-complete Java code integrated with legal prose. Consequently, code sharing and interaction for automated contract execution can be impractical between different blockchain platforms\footnote{\url{https://widgets.weforum.org/blockchain-toolkit/interoperability/index.html\#q01}, accessed Dec.\ 6, 2023.}. %Smart contracts are one of the key technologies in blockchain-based transactive energy systems to perform peer-to-peer energy trading, demand-response and flexibility trading, distributed control, etc. \cite{kirli2022smart}. However, 
The absence of standardization in smart contract design further complicates integration of transactive energy applications. Heterogeneous execution environments, like Ethereum Virtual Machine (EVM)\footnote{\url{https://ethereum.org/en/developers/docs/evm/}, accessed Dec. 6, 2023.} and Solana Virtual Machine (SVM)\footnote{\url{https://docs.solana.com}, accessed Dec. 6, 2023.}, have made it difficult to develop virtual machine-agnostic smart contracts.

\subsubsection{Concluding remarks}
Realizing seamless interoperability in blockchain-based energy trading necessitates addressing various challenges. Such challenges stem from the diversity of consensus mechanisms, lack of standardisation, and fragmented smart contract programming and execution environments. Overcoming these interoperability issues demands a collaborative effort from standardization bodies, regulatory authorities, and blockchain technology developers to establish common and efficient standards and ensure compliance.

\subsection{Blockchain scalability and throughput}

\subsubsection{Challenges in scalability}
The blockchain trilemma~\cite{dunphy2022note} outlines the inherent trade-offs between decentralisation (\ie distributed control and decision-making among network participants), security (\ie protection against unauthorized data processing), and scalability (\ie handling increasing transactions without sacrificing performance) in blockchain technology. For instance, in a proof-of-work based blockchains, increasing hash power in mining enhances network resistance to attack and improves security, but it reduces scalability and decentralization due to additional computation and communication resources. In contrast, reducing the number of miners can improve scalability by faster transaction processing, but it compromises security and decentralization since less miners may degrade the system robust and centralize the network. Thus, achieving high levels of decentralisation and security often comes at the expense of limited scalability, presenting a fundamental challenge in blockchain design.
%The blockchain trilemma~\cite{dunphy2022note} outlines the inherent trade-offs between decentralisation, security, and scalability in blockchain technology. \Ie decentralisation ensures no single entity control over the network, security protects against fraud, and scalability allows blockchains to handle a growing number of transactions and users while maintaining transaction throughput. Achieving high levels of decentralisation and security often comes at the expense of limited scalability, presenting a fundamental challenge in blockchain design.

The scalability concern in public blockchain is three-fold: transactions throughput, storage, and networking~\cite{xie2019survey}. Current blockchain throughput, exemplified by Bitcoin's seven transactions per second, falls far behind conventional payment systems like VISA's 2000 transactions per second. Enhancing throughput requires careful consideration on transaction volume, block interval time, and block size limitations. Particularly, higher transaction volume requires more frequent block creation. While decreasing block interval might increase block creation frequency, it may not be adequate since high transaction volume might also demand proportionally large block sizes~\cite{croman2016scaling}. % COMMENT: the last sentence does not clearly explain "why"
Regarding storage, the integration of blockchain in transactive energy systems necessitates processing substantial data from diverse devices, posing challenges for nodes with limited storage and computing resources. Furthermore, networking complexities arise as the traditional broadcast-centric data transmission mode can be inadequate for handling a large number of transactions.

\subsubsection{Potential solutions}
Solutions for scalable blockchain have been proposed in addressing the discussed scalability challenges, \eg Segregated Witness (SegWit), off-chain transactions, sharding, and Bitcoin-NG~\cite{xie2019survey}. The upgraded protocal for Bitcoin called SegWit~\cite{perez2019analysis} enhances throughput and maintains compatibility with existing infrastructure, yet with limited improvement on throughput. Off-chain transactions~\cite{decker2015fast} reduce on-chain transaction by storing them outside of the blockchain, but they compromise security and user experience due to the requirement of additional interface. Sharding~\cite{luu2016secure} is the technique that divides the blockchain into smaller partitions to improve throughput and reduce node load, while it sacrifices global consensus and introduces inter-shard transaction complexity. Bitcoin-NG~\cite{eyal2016bitcoin} is a Bitcoin variant with improved throughput, yet it risks double-spending attacks. 

\subsubsection{Concluding remarks}
There exists a trilemma in blockchain that intersects decentralisation, security, and scalability issues, and it poses significant challenges in the context of transactive energy systems. Integrating blockchain into transactive energy systems necessitates processing large amounts of data from diverse energy assets, thus straining the resources of nodes on the network. While various scalability solutions have been proposed, they often come with costs, such as reduced security or increased complexity. Achieving scalability while maintaining decentralization and security remains challenging for blockchain-enabled transactive energy applications. As blockchain adoption grows in the energy sector, innovative solutions will be essential to meet the growing demand for scalable and efficient blockchain systems.

\section{Breakdown of major %big
energy transition goals and pathway for this transition}
% Breakdown as a noun is one word.

This section looks into the energy transition where our transactive energy research takes background in. We analyze the feasibility of and barriers in achieving the energy transition, and suggest possible ways that can contribute to a clear pathway to this transition. 

Energy goals have been set to implement the energy transition, \eg the EU 2020 goal of 20\% greenhouse gas emission reduction and 20\% share of renewable energy~\cite{cross2015progress}, the EU 2030 goal with 55\% cuts in greenhouse gas emission and 32\% share of renewables~\cite{jager2020photovoltaics}, and the EU 2050 goal~\cite{wolf2021european} to be climate-neutral. Similar energy goals exist in Africa\footnote{\url{https://africandchub.org}, accessed Dec.\ 6, 2023.}\footnote{\url{https://unfccc.int/sites/default/files/NDC/2022-06/South\%20Africa\%20updated\%20first\%20NDC\%20September\%202021.pdf}, accessed Dec.\ 6, 2023.}~\cite{irena2015africa}, America\footnote{\url{https://www.whitehouse.gov/briefing-room/statements-releases/2021/01/27/fact-sheet-president-biden-takes-executive-actions-to-tackle-the-climate-crisis-at-home-and-abroad-create-jobs-and-restore-scientific-integrity-across-federal-government/}, accessed Dec.\ 6, 2023.}, Australia\footnote{\url{https://www.energy.gov.au/government-priorities/australias-energy-strategies-and-frameworks/national-energy-performance-strategy}, accessed Dec.\ 6, 2023.}, Brazil\footnote{\url{https://unfccc.int/sites/default/files/NDC/2022-06/Updated\%20-\%20First\%20NDC\%20-\%20\%20FINAL\%20-\%20PDF.pdf}, accessed Dec.\ 6, 2023.}\footnote{\url{https://www4.unfccc.int/sites/submissions/INDC/Published\%20Documents/Brazil/1/BRAZIL\%20iNDC\%20english\%20FINAL.pdf}, accessed Dec.\ 6, 2023.}\footnote{\url{https://www.europarl.europa.eu/RegData/etudes/BRIE/2022/738185/EPRS_BRI(2022)738185_EN.pdf}, accessed Dec.\ 6, 2023.}
, Canada\footnote{\url{https://www.canada.ca/en/services/environment/weather/climatechange/pan-canadian-framework/fourth-annual-report.html}, accessed Dec.\ 6, 2023.}, China\footnote{\url{https://unfccc.int/sites/default/files/resource/China\%E2\%80\%99s\%20Mid-Century\%20Long-Term\%20Low\%20Greenhouse\%20Gas\%20Emission\%20Development\%20Strategy.pdf}, accessed Dec.\ 6, 2023.}\footnote{\url{https://unfccc.int/sites/default/files/NDC/2022-06/China\%27s\%20First\%20NDC\%20Submission.pdf}, accessed Dec.\ 6, 2023.}, India\footnote{\url{https://unfccc.int/sites/default/files/NDC/2022-08/India\%20Updated\%20First\%20Nationally\%20Determined\%20Contrib.pdf}, accessed Dec.\ 6, 2023.}, Japan\footnote{\url{https://www.eu-japan.eu/news/japans-new-basic-energy-plan-until-2030-approved}, accessed Dec.\ 6, 2023.}\footnote{\url{https://www.europarl.europa.eu/RegData/etudes/BRIE/2021/698023/EPRS_BRI(2021)698023_EN.pdf}, accessed Dec.\ 6, 2023.}. EEA countries like Germany (see the National Energy Efficiency Action Plan\footnote{\url{https://www.energypartnership.cn/fileadmin/user_upload/china/media_elements/Documents/200407_BMWi_Dossier_Energy_Efficiency_Strategy_2050.pdf}, accessed Dec.\ 6, 2023.}) and Norway have broken down the envisaged cut of 55\% into sectors such as transport, building, industry, and committed the sectors to achieve the yearly reduction necessary to report on the yearly cut to reach the goals. 

However, reality shows that the envisaged yearly cut might not be achievable without major economical and societal costs. A study performed by TØI involving the main actors in the transport sector in Norway, including actors in public transport, train, ship, air travel shows that reaching the envisaged cut of 55\% is difficult, either through the measure of strong price increases for transport or assumed technology development and bio-blending\footnote{\url{https://www.toi.no/getfile.php?mmfileid=75433}, accessed Dec.\ 6, 2023.}. The 2020 EU-wide assessment on energy plans indicates a 2.8\% gap\footnote{\url{https://eur-lex.europa.eu/legal-content/EN/TXT/PDF/?uri=CELEX:52020DC0564\&from=EN}, accessed Dec.\ 6, 2023.} in primary energy consumption compared to the EU’s 2030 target of at least 32.5\%. Furthermore, though tremendous efforts have been made to break down energy goals\footnote{\url{https://energy.ec.europa.eu/topics/energy-strategy/national-energy-and-climate-plans-necps_en\#draft-necps}, accessed Dec.\ 6, 2023.}\footnote{\url{https://commission.europa.eu/energy-climate-change-environment/implementation-eu-countries/energy-and-climate-governance-and-reporting/national-energy-and-climate-plans_en\#final-necps}, accessed Dec.\ 6, 2023.}\footnote{\url{https://unfccc.int/NDCREG}, accessed Dec.\ 6, 2023.}\footnote{\url{https://eur-lex.europa.eu/legal-content/EN/TXT/PDF/?uri=CELEX:32018R0842\&from=EN}, accessed Dec.\ 6, 2023.} and the corresponding measures have been assessed at national and regional level (see the individual\footnote{\url{https://energy.ec.europa.eu/publications/individual-assessments_en}, accessed Dec.\ 6, 2023.} and EU-wide assessment), such experience has not been formalized and general guidelines in practice have not been established. It is also unclear whether big energy goals can be achieved in a disaggregated way, either geographically or categorically, in specific regions considering their diverse energy status and energy interactions in between them. Beyond that, there is limited knowledge about how future energy activities, \eg P2P or other transactive energy trading, might contribute to regional energy transition goals quantitatively~\cite{espadinha2023assessing, liu2022uncertainty,wu2022contribution}.

Closing the gap requires - not limited to - (i) the real-life oriented pathway for the electrical transition and (ii) modeling the dynamics amongst relevant roles and defining/quantifying contribution factors to the holistic and disaggregated energy goals. It is also critical to determine how a holistic goal can be partitioned into sub-goals for local regions and contribution share of specific energy sources, \eg solar or wind power. In this way, the achievement of energy goals can be visualized and calibrated in time to get pertinent feedback, leading to a more understandable and scrutinized energy transition. 

\section{Transactive energy management as transdisciplinary research}
Energy research has brought together various disciplines such as engineering, physics, environmental science, economics, psychology, and political science. This interdisciplinary approach is necessary because energy challenges come from different angles, \eg technical, environmental, economic, and social dimensions. %For instance, the development of renewable energy technologies involves not only engineering and physics but also an understanding of environmental impacts, economic and social viability, and adapting existing policy frameworks. The University of Oslo’s Privacy Preserving Transactive Energy Management (PriTEM) exemplifies such transdisciplinary research. Designed to develop theoretical and practical knowledge on how transactive energy management can contribute to energy decentralization and democratization of energy management, PriTEM is organized as as a transdisciplinary project consisting of researchers from four departments, namely Department of Informatics, Faculty of Law, Department of Technology Systems, and Department of Psychology. The intention of having a transdisciplinary project team is rooted in the recognition that transactive energy management is a complex problem that require insights and approaches from multiple disciplines.
Identifying and solving the issues in the transdisciplinary studies on transactive energy is complex that requires insights and approaches from multiple disciplines. In that way, it helps researchers to transcend boundaries across research fields and foster holistic and problem-solving methodologies.

%The works of the researchers of these four faculties remain under the management of the respective faculty. The outputs are consequently classified into different Work Packages. Ideally, the output of one Work Package should constitute partial input for the work of the subsequent Work Package. The output of WP1 should contribute to the research work in WP2 and the knowledge produced in WP2 should enable further study in WP3, and so on. While the structure of the collaboration seems clear and logical, putting this in practice is far more challenging. 

Drawing on our experience on transactive energy studies where we collaborate across disciplines, several challenges in this transdisciplinary collaboration include, but are not limited to: 
(i) the difficulty in setting a common language, as field-specific jargons become inevitable when discussion evolves in depth; (ii) the need to distinguish between issues that can be solved without collaboration and those that cannot; (iii) the ability to identify the granularity of the discussion topic that ensures the relevance of the discussion to the attendants; (iv) the need to balance between individual priorities and collective goals of the team; % COMMENT: the same with (iv). (5) dealing with the elusive end-goal, whether we are collaborating to produce something together or to produce something individually based on the inputs from the others; 
(v) the ability to admit that ``I do not know'' in unfamiliar topics and seek help from others to fill the knowledge gap.

While it remains challenging and no fixed strategy to practice cross-disciplinary collaboration, we have learned some valuable lessons from our collaboration work. The most important is, perhaps, to foster a conducive and trusting climate for collaboration where members feel safe to ask questions and voice concerns. It can also be useful to set concrete and achievable milestones that members can work towards together. The collaboration so far has shown that we are heading in the right direction. After the initial phase of building a common terminology base in our collaboration, we envisage to establish the framework on ``how to learn from each other''.

\section*{Acknowledgment}

This work was supported via the grant ``Privacy preserving Transactive Energy Management (PriTEM)'' funded by UiO:Energy Convergence Environments.

\bibliographystyle{IEEEtran}
\bibliography{references}

\begin{thebibliography}{10}
\providecommand{\url}[1]{#1}
\csname url@samestyle\endcsname
\providecommand{\newblock}{\relax}
\providecommand{\bibinfo}[2]{#2}
\providecommand{\BIBentrySTDinterwordspacing}{\spaceskip=0pt\relax}
\providecommand{\BIBentryALTinterwordstretchfactor}{4}
\providecommand{\BIBentryALTinterwordspacing}{\spaceskip=\fontdimen2\font plus
\BIBentryALTinterwordstretchfactor\fontdimen3\font minus
  \fontdimen4\font\relax}
\providecommand{\BIBforeignlanguage}[2]{{%
\expandafter\ifx\csname l@#1\endcsname\relax
\typeout{** WARNING: IEEEtran.bst: No hyphenation pattern has been}%
\typeout{** loaded for the language `#1'. Using the pattern for}%
\typeout{** the default language instead.}%
\else
\language=\csname l@#1\endcsname
\fi
#2}}
\providecommand{\BIBdecl}{\relax}
\BIBdecl

\bibitem{osti_1123244}
\BIBentryALTinterwordspacing
R.~B. Melton, ``Gridwise transactive energy framework (draft version),'' Nov.
  2013. [Online]. Available: \url{https://www.osti.gov/biblio/1123244}
\BIBentrySTDinterwordspacing

\bibitem{EES-031}
\BIBentryALTinterwordspacing
W.~Tushar, S.~Nizami, M.~I. Azim, C.~Yuen, D.~B. Smith, T.~Saha, and H.~V.
  Poor, ``Peer-to-peer energy sharing: A comprehensive review,''
  \emph{Foundations and Trends® in Electric Energy Systems}, vol.~6, no.~1,
  pp. 1--82, 2023. [Online]. Available:
  \url{http://dx.doi.org/10.1561/3100000031}
\BIBentrySTDinterwordspacing

\bibitem{8582548}
\BIBentryALTinterwordspacing
S.~Wu, F.~Zhang, and D.~Li, ``User-centric peer-to-peer energy trading
  mechanisms for residential microgrids,'' in \emph{2018 2nd IEEE Conference on
  Energy Internet and Energy System Integration (EI2)}, 2018, pp. 1--6.
  [Online]. Available: \url{https://doi.org/10.1109/EI2.2018.8582548}
\BIBentrySTDinterwordspacing

\bibitem{10.12688/openreseurope.15282.1}
\BIBentryALTinterwordspacing
I.~Pigliautile, S.~Breukers, M.~Boekelo, P.~Carnero, F.~Causone, S.~Arko,
  S.~Ferroni, B.~Pioppi, A.~Pisello, A.~Solar, J.~Swens, E.~Tarpani, and
  S.~D'Oca, ``Peer-to-peer energy communities: regulatory barriers in the eu
  context [version 1; peer review: 4 approved with reservations],'' \emph{Open
  Research Europe}, vol.~2, no. 147, 2022. [Online]. Available:
  \url{https://doi.org/10.12688/openreseurope.15282.1}
\BIBentrySTDinterwordspacing

\bibitem{ZHOU2020739}
\BIBentryALTinterwordspacing
Y.~Zhou, J.~Wu, C.~Long, and W.~Ming, ``State-of-the-art analysis and
  perspectives for peer-to-peer energy trading,'' \emph{Engineering}, vol.~6,
  no.~7, pp. 739--753, 2020. [Online]. Available:
  \url{https://doi.org/10.1016/j.eng.2020.06.002}
\BIBentrySTDinterwordspacing

\bibitem{proceedings2231472}
\BIBentryALTinterwordspacing
F.~Bandeiras, M.~Gomes, P.~Coelho, and J.~Fernandes, ``Net zero energy for
  industrial and commercial microgrids: Approaches and challenges,''
  \emph{Proceedings}, vol.~2, no.~23, 2018. [Online]. Available:
  \url{https://www.mdpi.com/2504-3900/2/23/1472}
\BIBentrySTDinterwordspacing

\bibitem{Chapter12Longtermpowerpurchaseagreementsthefactorsthatinfluencecontractdesign}
\BIBentryALTinterwordspacing
P.~Wallace, \emph{Chapter 12: Long-term power purchase agreements: the factors
  that influence contract design}.\hskip 1em plus 0.5em minus 0.4em\relax
  Cheltenham, UK: Edward Elgar Publishing, 2019. [Online]. Available:
  \url{https://www.elgaronline.com/view/edcoll/9781786436146/9781786436146.00023.xml}
\BIBentrySTDinterwordspacing

\bibitem{POULLIKKAS20131}
\BIBentryALTinterwordspacing
A.~Poullikkas, ``A comparative assessment of net metering and feed in tariff
  schemes for residential pv systems,'' \emph{Sustainable Energy Technologies
  and Assessments}, vol.~3, pp. 1--8, 2013. [Online]. Available:
  \url{https://www.sciencedirect.com/science/article/pii/S2213138813000313}
\BIBentrySTDinterwordspacing

\bibitem{en15031229}
\BIBentryALTinterwordspacing
S.~Junlakarn, P.~Kokchang, and K.~Audomvongseree, ``Drivers and challenges of
  peer-to-peer energy trading development in thailand,'' \emph{Energies},
  vol.~15, no.~3, 2022. [Online]. Available:
  \url{https://www.mdpi.com/1996-1073/15/3/1229}
\BIBentrySTDinterwordspacing

\bibitem{MALDET2022131805}
\BIBentryALTinterwordspacing
M.~Maldet, F.~H. Revheim, D.~Schwabeneder, G.~Lettner, P.~C. {del Granado},
  A.~Saif, M.~Löschenbrand, and S.~Khadem, ``Trends in local electricity
  market design: Regulatory barriers and the role of grid tariffs,''
  \emph{Journal of Cleaner Production}, vol. 358, p. 131805, 2022. [Online].
  Available:
  \url{https://www.sciencedirect.com/science/article/pii/S0959652622014159}
\BIBentrySTDinterwordspacing

\bibitem{cavoukian2010privacy}
\BIBentryALTinterwordspacing
A.~Cavoukian, S.~Taylor, and M.~E. Abrams, ``Privacy by design: essential for
  organizational accountability and strong business practices,'' \emph{Identity
  in the Information Society}, vol.~3, pp. 405--413, 2010. [Online]. Available:
  \url{https://doi.org/10.1007/s12394-010-0053-z}
\BIBentrySTDinterwordspacing

\bibitem{koops2014privacy}
\BIBentryALTinterwordspacing
B.-J. Koops and R.~Leenes, ``Privacy regulation cannot be hardcoded. a critical
  comment on the ‘privacy by design’provision in data-protection law,''
  \emph{International Review of Law, Computers \& Technology}, vol.~28, no.~2,
  pp. 159--171, 2014. [Online]. Available:
  \url{https://doi.org/10.1080/13600869.2013.801589}
\BIBentrySTDinterwordspacing

\bibitem{soto2021peer}
\BIBentryALTinterwordspacing
E.~A. Soto, L.~B. Bosman, E.~Wollega, and W.~D. Leon-Salas, ``Peer-to-peer
  energy trading: A review of the literature,'' \emph{Applied Energy}, vol.
  283, p. 116268, 2021. [Online]. Available:
  \url{https://doi.org/10.1016/j.apenergy.2020.116268}
\BIBentrySTDinterwordspacing

\bibitem{dalen2017towards}
\BIBentryALTinterwordspacing
A.~Dal{\'e}n and J.~Kr{\"a}mer, ``Towards a user-centered feedback design for
  smart meter interfaces to support efficient energy-use choices: A design
  science approach,'' \emph{Business \& Information Systems Engineering},
  vol.~59, pp. 361--373, 2017. [Online]. Available:
  \url{https://doi.org/10.1007/s12599-017-0489-x}
\BIBentrySTDinterwordspacing

\bibitem{ahram2010user}
\BIBentryALTinterwordspacing
T.~Z. Ahram, W.~Karwowski, and B.~Amaba, ``User-centered systems engineering \&
  knowledge management framework for design \& modeling of future smart
  cities,'' in \emph{Proceedings of the Human Factors and Ergonomics Society
  Annual Meeting}, vol.~54, no.~20.\hskip 1em plus 0.5em minus 0.4em\relax SAGE
  Publications Sage CA: Los Angeles, CA, 2010, pp. 1752--1756. [Online].
  Available: \url{https://doi.org/10.1177/154193121005402002}
\BIBentrySTDinterwordspacing

\bibitem{savchukuser}
\BIBentryALTinterwordspacing
O.~Savchuk, H.~C. Moll, and J.~W. Turkstra, ``User-centered design and
  evaluation of decentralized energy systems in the netherlands.'' [Online].
  Available:
  \url{http://www.energy-proceedings.org/wp-content/uploads/2020/11/aeab2020_paper_181.pdf}
\BIBentrySTDinterwordspacing

\bibitem{ruotsalainen2017culture}
\BIBentryALTinterwordspacing
J.~Ruotsalainen, J.~Karjalainen, M.~Child, and S.~Heinonen, ``Culture, values,
  lifestyles, and power in energy futures: A critical peer-to-peer vision for
  renewable energy,'' \emph{Energy Research \& Social Science}, vol.~34, pp.
  231--239, 2017. [Online]. Available:
  \url{https://doi.org/10.1016/j.erss.2017.08.001}
\BIBentrySTDinterwordspacing

\bibitem{mengelkamp2018quantifying}
\BIBentryALTinterwordspacing
E.~Mengelkamp, P.~Staudt, J.~G{\"a}rttner, C.~Weinhardt, and J.~Huber,
  ``Quantifying factors for participation in local electricity markets,'' in
  \emph{2018 15th International Conference on the European Energy Market
  (EEM)}.\hskip 1em plus 0.5em minus 0.4em\relax IEEE, 2018, pp. 1--5.
  [Online]. Available: \url{https://doi.org/10.1109/EEM.2018.8469969}
\BIBentrySTDinterwordspacing

\bibitem{georgarakis2021keep}
\BIBentryALTinterwordspacing
E.~Georgarakis, T.~Bauwens, A.-M. Pronk, and T.~AlSkaif, ``Keep it green,
  simple and socially fair: A choice experiment on prosumers’ preferences for
  peer-to-peer electricity trading in the netherlands,'' \emph{Energy Policy},
  vol. 159, p. 112615, 2021. [Online]. Available:
  \url{https://doi.org/10.1016/j.enpol.2021.112615}
\BIBentrySTDinterwordspacing

\bibitem{li2022preference}
\BIBentryALTinterwordspacing
D.~Li, J.-H. Bae, and M.~Rishi, ``A preference analysis for a peer-to-peer
  (p2p) electricity trading platform in south korea,'' \emph{Energies},
  vol.~15, no.~21, p. 7973, 2022. [Online]. Available:
  \url{https://doi.org/10.3390/en15217973}
\BIBentrySTDinterwordspacing

\bibitem{hackbarth2020attitudes}
\BIBentryALTinterwordspacing
A.~Hackbarth and S.~L{\"o}bbe, ``Attitudes, preferences, and intentions of
  german households concerning participation in peer-to-peer electricity
  trading,'' \emph{Energy Policy}, vol. 138, p. 111238, 2020. [Online].
  Available: \url{https://doi.org/10.1016/j.enpol.2020.111238}
\BIBentrySTDinterwordspacing

\bibitem{karami2022business}
\BIBentryALTinterwordspacing
M.~Karami and R.~Madlener, ``Business models for peer-to-peer energy trading in
  germany based on households’ beliefs and preferences,'' \emph{Applied
  Energy}, vol. 306, p. 118053, 2022. [Online]. Available:
  \url{https://doi.org/10.1016/j.apenergy.2021.118053}
\BIBentrySTDinterwordspacing

\bibitem{huang2021review}
\BIBentryALTinterwordspacing
Q.~Huang, W.~Amin, K.~Umer, H.~B. Gooi, F.~Y.~S. Eddy, M.~Afzal, M.~Shahzadi,
  A.~A. Khan, and S.~A. Ahmad, ``A review of transactive energy systems:
  Concept and implementation,'' \emph{Energy Reports}, vol.~7, pp. 7804--7824,
  2021. [Online]. Available: \url{https://doi.org/10.1016/j.egyr.2021.05.037}
\BIBentrySTDinterwordspacing

\bibitem{andoni2019blockchain}
\BIBentryALTinterwordspacing
M.~Andoni, V.~Robu, D.~Flynn, S.~Abram, D.~Geach, D.~Jenkins, P.~McCallum, and
  A.~Peacock, ``Blockchain technology in the energy sector: A systematic review
  of challenges and opportunities,'' \emph{Renewable and sustainable energy
  reviews}, vol. 100, pp. 143--174, 2019. [Online]. Available:
  \url{https://doi.org/10.1016/j.rser.2018.10.014}
\BIBentrySTDinterwordspacing

\bibitem{nakamoto2008bitcoin}
\BIBentryALTinterwordspacing
S.~Nakamoto, ``Bitcoin: A peer-to-peer electronic cash system,'' 2008.
  [Online]. Available: \url{https://bitcoin.org/bitcoin.pdf}
\BIBentrySTDinterwordspacing

\bibitem{narayanan2016bitcoin}
\BIBentryALTinterwordspacing
A.~Narayanan, J.~Bonneau, E.~Felten, A.~Miller, and S.~Goldfeder,
  \emph{{Bitcoin and Cryptocurrency Technologies: A Comprehensive
  Introduction}}.\hskip 1em plus 0.5em minus 0.4em\relax Princeton University
  Press, 2016. [Online]. Available:
  \url{http://press.princeton.edu/titles/10908.html}
\BIBentrySTDinterwordspacing

\bibitem{szabo1996smart}
N.~Szabo, ``Smart contracts: building blocks for digital markets,''
  \emph{EXTROPY: The Journal of Transhumanist Thought,(16)}, vol.~18, no.~2,
  p.~28, 1996.

\bibitem{kirli2022smart}
\BIBentryALTinterwordspacing
D.~Kirli, B.~Couraud, V.~Robu, M.~Salgado-Bravo, S.~Norbu, M.~Andoni,
  I.~Antonopoulos, M.~Negrete-Pincetic, D.~Flynn, and A.~Kiprakis, ``Smart
  contracts in energy systems: A systematic review of fundamental approaches
  and implementations,'' \emph{Renewable and Sustainable Energy Reviews}, vol.
  158, p. 112013, 2022. [Online]. Available:
  \url{https://doi.org/10.1016/j.rser.2021.112013}
\BIBentrySTDinterwordspacing

\bibitem{Dubey2022blockchain}
``How does blockchain technology ensure transparency in cryptocurrency trade,''
  \url{https://hashstudioz.com/blog/how-does-blockchain-technology-ensure-transparency-in-cryptocurrency-trade/},
  accessed: 2023-11-24.

\bibitem{peng2021privacy}
\BIBentryALTinterwordspacing
L.~Peng, W.~Feng, Z.~Yan, Y.~Li, X.~Zhou, and S.~Shimizu, ``Privacy
  preservation in permissionless blockchain: A survey,'' \emph{Digital
  Communications and Networks}, vol.~7, no.~3, pp. 295--307, 2021. [Online].
  Available: \url{https://doi.org/10.1016/j.dcan.2020.05.008}
\BIBentrySTDinterwordspacing

\bibitem{Chantal2022privacy}
``The privacy paradox in blockchain: best practices for data management in
  crypto,''
  \url{https://www.dentons.com/en/insights/articles/2022/june/9/the-privacy-paradox-in-blockchain-best-practices-for-data-management-in-crypto},
  accessed: 2023-11-24.

\bibitem{conklin2004principles}
\BIBentryALTinterwordspacing
A.~Conklin, G.~White, C.~Cothren, D.~Williams, and R.~L. Davis,
  \emph{Principles of computer security: security+ and beyond}.\hskip 1em plus
  0.5em minus 0.4em\relax McGraw-Hill, Inc., 2004. [Online]. Available:
  \url{https://doi.org/10.1007/s10489-010-0263-y}
\BIBentrySTDinterwordspacing

\bibitem{barsan2019public}
\BIBentryALTinterwordspacing
I.~M. Barsan, ``Public blockchains: the privacy-transparency conundrum,''
  \emph{Revue Trimestrielle de Droit Financier (RTDF) N}, pp. 2--2019, 2019.
  [Online]. Available: \url{https://ssrn.com/abstract=3445025}
\BIBentrySTDinterwordspacing

\bibitem{dunphy2022note}
\BIBentryALTinterwordspacing
P.~Dunphy, ``A note on the blockchain trilemma for decentralized identity:
  Learning from experiments with hyperledger indy,'' \emph{arXiv preprint
  arXiv:2204.05784}, 2022. [Online]. Available:
  \url{https://doi.org/10.48550/arXiv.2204.05784}
\BIBentrySTDinterwordspacing

\bibitem{xie2019survey}
\BIBentryALTinterwordspacing
J.~Xie, F.~R. Yu, T.~Huang, R.~Xie, J.~Liu, and Y.~Liu, ``A survey on the
  scalability of blockchain systems,'' \emph{IEEE network}, vol.~33, no.~5, pp.
  166--173, 2019. [Online]. Available:
  \url{https://doi.org/10.1109/MNET.001.1800290}
\BIBentrySTDinterwordspacing

\bibitem{croman2016scaling}
\BIBentryALTinterwordspacing
K.~Croman, C.~Decker, I.~Eyal, A.~E. Gencer, A.~Juels, A.~Kosba, A.~Miller,
  P.~Saxena, E.~Shi, E.~G{\"u}n~Sirer \emph{et~al.}, ``On scaling decentralized
  blockchains: (a position paper),'' in \emph{International conference on
  financial cryptography and data security}.\hskip 1em plus 0.5em minus
  0.4em\relax Springer, 2016, pp. 106--125. [Online]. Available:
  \url{https://doi.org/10.1007/978-3-662-53357-4_8}
\BIBentrySTDinterwordspacing

\bibitem{perez2019analysis}
\BIBentryALTinterwordspacing
C.~P{\'e}rez-Sol{\`a}, S.~Delgado-Segura, J.~Herrera-Joancomart{\i}, and
  G.~Navarro-Arribas, ``Analysis of the segwit adoption in bitcoin,'' 2019.
  [Online]. Available:
  \url{https://deic.uab.cat/~guille/files/papers/2018.recsi.segwit.pdf}
\BIBentrySTDinterwordspacing

\bibitem{decker2015fast}
\BIBentryALTinterwordspacing
C.~Decker and R.~Wattenhofer, ``A fast and scalable payment network with
  bitcoin duplex micropayment channels,'' in \emph{Stabilization, Safety, and
  Security of Distributed Systems: 17th International Symposium, SSS 2015,
  Edmonton, AB, Canada, August 18-21, 2015, Proceedings 17}.\hskip 1em plus
  0.5em minus 0.4em\relax Springer, 2015, pp. 3--18. [Online]. Available:
  \url{https://doi.org/10.1007/978-3-319-21741-3_1}
\BIBentrySTDinterwordspacing

\bibitem{luu2016secure}
\BIBentryALTinterwordspacing
L.~Luu, V.~Narayanan, C.~Zheng, K.~Baweja, S.~Gilbert, and P.~Saxena, ``A
  secure sharding protocol for open blockchains,'' in \emph{Proceedings of the
  2016 ACM SIGSAC conference on computer and communications security}, 2016,
  pp. 17--30. [Online]. Available:
  \url{https://doi.org/10.1145/2976749.2978389}
\BIBentrySTDinterwordspacing

\bibitem{eyal2016bitcoin}
\BIBentryALTinterwordspacing
I.~Eyal, A.~E. Gencer, E.~G. Sirer, and R.~Van~Renesse, ``$\{$Bitcoin-NG$\}$: A
  scalable blockchain protocol,'' in \emph{13th USENIX symposium on networked
  systems design and implementation (NSDI 16)}, 2016, pp. 45--59. [Online].
  Available:
  \url{https://www.usenix.org/conference/nsdi16/technical-sessions/presentation/eyal}
\BIBentrySTDinterwordspacing

\bibitem{cross2015progress}
\BIBentryALTinterwordspacing
S.~Cross, A.~Hast, R.~Kuhi-Thalfeldt, S.~Syri, D.~Streimikiene, and A.~Denina,
  ``Progress in renewable electricity in northern europe towards eu 2020
  targets,'' \emph{Renewable and Sustainable Energy Reviews}, vol.~52, pp.
  1768--1780, 2015. [Online]. Available:
  \url{https://doi.org/10.1016/j.rser.2015.07.165}
\BIBentrySTDinterwordspacing

\bibitem{jager2020photovoltaics}
\BIBentryALTinterwordspacing
A.~J{\"a}ger-Waldau, I.~Kougias, N.~Taylor, and C.~Thiel, ``How photovoltaics
  can contribute to ghg emission reductions of 55\% in the eu by 2030,''
  \emph{Renewable and Sustainable Energy Reviews}, vol. 126, p. 109836, 2020.
  [Online]. Available: \url{https://doi.org/10.1016/j.rser.2020.109836}
\BIBentrySTDinterwordspacing

\bibitem{wolf2021european}
\BIBentryALTinterwordspacing
S.~Wolf, J.~Teitge, J.~Mielke, F.~Sch{\"u}tze, and C.~Jaeger, ``The european
  green deal—more than climate neutrality,'' \emph{Intereconomics}, vol.~56,
  pp. 99--107, 2021. [Online]. Available:
  \url{https://doi.org/10.1007/s10272-021-0963-z}
\BIBentrySTDinterwordspacing

\bibitem{irena2015africa}
\BIBentryALTinterwordspacing
I.~IRENA, ``Africa 2030: Roadmap for a renewable energy future,'' \emph{IRENA:
  Abu Dhabi, Saudi Arabia}, 2015. [Online]. Available:
  \url{https://www.irena.org/-/media/Files/IRENA/Agency/Publication/2015/IRENA_Africa_2030_REmap_2015_low-res.pdf}
\BIBentrySTDinterwordspacing

\bibitem{espadinha2023assessing}
\BIBentryALTinterwordspacing
J.~Espadinha, P.~Baptista, and D.~Neves, ``Assessing p2p energy markets
  contribution for 2050 decarbonization goals,'' \emph{Sustainable Cities and
  Society}, vol.~92, p. 104495, 2023. [Online]. Available:
  \url{https://doi.org/10.1016/j.scs.2023.104495}
\BIBentrySTDinterwordspacing

\bibitem{liu2022uncertainty}
\BIBentryALTinterwordspacing
J.~Liu, Y.~Zhou, H.~Yang, and H.~Wu, ``Uncertainty energy planning of net-zero
  energy communities with peer-to-peer energy trading and green vehicle storage
  considering climate changes by 2050 with machine learning methods,''
  \emph{Applied Energy}, vol. 321, p. 119394, 2022. [Online]. Available:
  \url{https://doi.org/10.1016/j.apenergy.2022.119394}
\BIBentrySTDinterwordspacing

\bibitem{wu2022contribution}
\BIBentryALTinterwordspacing
T.~Wu, S.~Wang, L.~Wang, and X.~Tang, ``Contribution of china's online
  car-hailing services to its 2050 carbon target: Energy consumption assessment
  based on the gcam-se model,'' \emph{Energy Policy}, vol. 160, p. 112714,
  2022. [Online]. Available: \url{https://doi.org/10.1016/j.enpol.2021.112714}
\BIBentrySTDinterwordspacing

\end{thebibliography}

\end{document}